\documentclass[pra,aps,preprint,onecolumn]{revtex4}

\usepackage[12pt]{xy}
\usepackage{color}
\usepackage{graphics}
\usepackage{epsfig}
\usepackage{amsfonts}
\usepackage{bm}

\begin{document}

\title{Separability of entangled qutrits in noisy channels}

\author{Agata Ch\c{e}ci\'nska} \email{agata.checinska@fuw.edu.pl}
\affiliation{Instytut Fizyki Teoretycznej, Uniwersytet Warszawski,
  Warszawa 00--681, Poland}
\affiliation{ICFO-Institut de Ciencies Fotoniques, Mediterranean Technology Park, 
08860 Castelldefels (Barcelona), Spain}

\author{Krzysztof W\'odkiewicz}
\email{wodkiew@fuw.edu.pl} \affiliation{Instytut Fizyki Teoretycznej,
  Uniwersytet Warszawski, Warszawa 00--681, Poland}

\affiliation{Department of Physics and Astronomy, University of New
  Mexico, Albuquerque, NM~87131-1156, USA}
\date{\today}

\begin{abstract}
We present an analysis of noisy atomic channels involving qutrits.  We choose a three-level atom with
V-configuration to be the qutrit state. Gell-Mann matrices and a generalized Bloch vector (8-dimensional)
are used to describe the qutrit density operator. We  introduce quantum quasi-distributions for qutrits that
provide a simple description of  entanglement. Studying the time-evolution for the atomic variables we find
the Kraus representation of spontaneous emission quantum channel (SE channel). Furthermore, we consider a
generalized Werner state of two qutrits  and investigate the separability condition in the presence of
spontaneous emission noise.  The influence of spontaneous emission on the separability of Werner states for
qutrit and qubit states is compared.
\end{abstract}

\maketitle

\section{Introduction}

Quantum information can be stored, transmitted and retrieved using light, cold ions or atoms.  Only in
highly ideal conditions these physical systems can be regarded as isolated and immune to various sources of
decoherence. Quantum channels based on atoms or photons are examples of open quantum systems interacting
with an environment, causing degradation of  the linear superpositions or the quantum non-separability of
correlated systems. The understanding and the control of such noisy channels is at the core of quantum
communication.

Experimental teleportation of atoms  \cite{blatt,wineland} provides an example of a
 channel   in which quantum information is  transferred with a high
fidelity.  The quantum teleportation protocol uses as a resource entangled atoms.  Entanglement of atoms can
be achieved using different physical phenomena such as coherent cold collisions \cite{zoller} or an optical
lattice \cite{hansch}. Recent experimental and theoretical investigations have shown that cold atoms and
individual photons may lead the way towards chip-scale quantum information processors \cite{monroe}.

One of the physical processes that may deteriorate the efficiency of atomic applicability, is spontaneous
emission. Dissipation induced by vacuum fluctuations in quantum channels with atoms, impacts atomic
entanglement and the fidelity of quantum protocols based on atomic systems.

In most  atomic applications to atomic channels,  the main building blocks of information was based on
two-level quantum systems, or qubits.  Using  N-level systems, or qudits, can in principle  improve the
efficiency of quantum channel due to a larger Hilbert spaces. It is known that entangled qudits can provide
a higher degree of efficiency  in quantum protocols \cite{spekkens}.

The simplest generalization of the qubit involves a qutrit i.e., a quantum state spanned by three
orthonormal states $|1\rangle, |2\rangle$ and $|3\rangle$. Qutrits can be physically implemented using three
level atoms \cite{jin}, transverse spatial modes of single photons \cite{langford}, or polarization
states of a single-mode biphoton field \cite{bogdanov}.

The goal of this paper is to discuss the properties of noisy atomic channels involving qutrits. The physical
realization of the qutrit state in a noisy channel will be based on a three-level atom with spontaneous
emission. Qutrit quantum channels with vacuum fluctuations are open quantum systems. It is the purpose of
this paper to study the properties of such noisy quantum channels.
 We shall investigate the efficiency, the fidelity of such  channels and their  impact on
the quantum separability on entangled  qutrits.

The paper is structured as follows. In  Section 2 we review the Bloch description of qutrits based on the
$SU(3)$ generators. We introduce and investigate quantum quasi-distributions for qutrits that provide a
simple description of qutrit  entanglement. We explain why a Werner mixed qutrit state is more robust
compared to
the qubit situation.\\
 In Section 3 we discuss spontaneous emission in the framework of
 the qutrit Bloch formalism, and  derive the Kraus representation for a qutrit noisy channel.\\
 In Section 4 we examine the influence of SE channel on state separability. We investigate when the impact of this noisy channel is stronger for qubits than for qutrits. The fidelity of the channel is computed and compared.\\
In section 5 we present a concise summary of our results.

\section{From qubit to qutrit states}

It is well known that a qubit  -- a quantum state living in a two dimensional Hilbert space, is
used as a basic building block of Quantum Information \cite{nielsen}. Within the framework of atomic physics
two-level atoms are  the simplest physical realizations  of  qubits \cite{eberly}. Many papers have been
written  on the subject of qubits and quantum  qubit channels \cite{nielsen,daffer,ruskai}. A natural
generalization of a qubit  to N-dimensional, involving qudits, has been investigated
\cite{kimura,hofman,cabello}, though has received less interest. From the physical point of view the use of
more complex atomic structures might be advantageous \cite{kaszlikowski}. The first natural  step in this
generalization brings us to quantum objects that belong to a three dimensional Hilbert space $\mathcal{H}^3$
- qutrits.

It is the purpose of this section to provide a useful description of qutrit states using such  tools like
the concept of the Bloch vector associated with a  Bloch sphere and  apply quantum quasi-distribution
functions for the description of qutrit states. We shall exemplify our approach discussing in a parallel way
qubit and qutrit properties.

\subsection{Qutrit Bloch vectors}

It is very advantageous to provide the mathematical description of qutrits in a similar way as qubits are
characterized with the use of the Bloch formalism. This formalism uses in an intrinsic way the $SU(2)$
generators, Pauli matrices, as a basis for the  qubit density operator

\begin{equation}
\rho_{qubit}=\frac{1}{2}(\mathbb{I}+\vec{n}\cdot\vec{\sigma})\,,
\end{equation}
where $\vec{n}=n_i \vec{e}_j $ is a three dimensional (real) Bloch vector. For a system of correlated qubits
$a$ and $b$, the corresponding density operator has the form

\begin{eqnarray}
    \rho^{ab}_{qubit}&=&\frac{1}{4}(\mathbb{I}_a \otimes \mathbb{I}_b +\vec{n}_a\cdot\vec{\sigma}\otimes \mathbb{I}_b +
     \mathbb{I}_a\otimes\vec{n}_b\cdot\vec{\sigma} +\nonumber \\ 
\  &\  &+\mathcal{C}_{ij}\sigma_i\otimes\sigma_j)\,.
\end{eqnarray}
where $\vec{n}_a$ and $\vec{n}_b$ are individual Bloch vectors of the two qubits and $\mathcal{C}_{ij} =
\langle \sigma_i\otimes \sigma_j \rangle$ is the correlation matrix  of the two  qubits.

For a maximally entangled state of the two qubits
\begin{equation}
|\Psi_{qubit}\rangle =\frac{1}{\sqrt{2}}(|1,1\rangle+|2,2\rangle)\,,
\end{equation}
written in the qubit basis $|1\rangle$ and  $|2\rangle$, the mean values of the individual Bloch vectors are
zero and the correlation matrix is diagonal and has the simple form
\begin{equation}
\label{corr2}
 \mathcal{C}_{ij}
=\textit{s}\delta_{ij}\,,
\end{equation}
where   $\textit{s} =(1,-1,1)$ corresponds to a sign assigned to the three corresponding components of the
Kronecker delta.

It is clear that the mathematical description of a qutrit density operator involves in a natural way the
$SU(3)$ generators, called the Gell-Mann matrices $\lambda_i$ \cite{gellmann}. Earlier applications of the
$SU(3)$ formalism to three level atoms can be found in references \cite{hioe}. More recent applications of
this formalism involving entanglement are presented in references \cite{caves}. The density oprator of the
qutrit is
\begin{equation}
\rho=\frac{1}{3}(\mathbb{I}+\sqrt{3}\;\vec{n}\cdot\vec{\lambda}),
\end{equation}
where $\vec{n}=n_i \vec{e}_i $ is now a real eight dimensional generalized Bloch vector. The Gell-Mann
matrices like the Pauli matrices are traceless and satisfy

$$\lambda_i \lambda_j =\frac{2}{3}\delta_{ij}+ d_{ijk}\lambda_k +i f_{ijk}
\lambda_k\,,$$ where the completely antisymmetric $f_{ijk}$ are the structure constants of the $SU(3)$
algebra, and $d_{ijk}$ are completely symmetric. Values of these coefficients and the explicit form of the
eight Gell-Mann matrices can be found in \cite{caves}.

Pure qutrit states correspond to vectors that satisfy
\begin{equation}
\vec{n}\cdot\vec{n}=1,\  \vec{n}*\vec{n}= \vec{e}_i d_{ijk}n_j n_k =\vec{n}\,.
\end{equation}
These two conditions define a generalized Bloch sphere for qutrits, in analogy to Bloch qubit sphere. Hence,
pure qutrit states in a unique way  refer to unit vectors $\vec{n}\in \mathcal{S}^7$, the seven-dimensional
unit sphere in $\mathcal{R}^8$ (first condition). However, the  second condition places three additional
constraints on the real parameters defining the pure state vector.

For a system of two correlated qutrits  $a$ and $b$, the corresponding density operator has the form

\begin{eqnarray}
    \rho_{qutrit}^{ab}&=&\frac{1}{9}(\mathbb{I}_a \otimes \mathbb{I}_b +\sqrt{3}\;\vec{n}_a\cdot\vec{\lambda}\otimes
    \mathbb{I}_b +\nonumber\\
 &+&\sqrt{3}\;\mathbb{I}_a\otimes\vec{n}_b\cdot\vec{\lambda} +3\;  \mathcal{C}_{ij}\lambda _i\otimes \lambda_j),.
\end{eqnarray}
where $\vec{n}_a$ and $\vec{n}_b$ are individual Bloch vectors of the two qutrits and $ \mathcal{C}_{ij} =
\frac{3}{4}\langle \lambda_i\otimes \lambda _j \rangle$ is the correlation matrix  of the two qutrits. The
maximally entangled state of the two qutrits is

\begin{equation}
   |\Psi\rangle =\frac{1}{\sqrt{3}}(|1,1\rangle+|2,2\rangle +
|3,3\rangle)\,,
\end{equation}
where a third state $|3\rangle$ has been added to the qubit maximally entangled state. In this case the mean
values of the individual Bloch vectors are zero and the $8\times8$ correlation matrix is diagonal

\begin{equation}
\label{corr3}
 \mathcal{C}_{ij}
=\frac{\textit{s}}{2}\delta_{ij}\,,
\end{equation}
where   $\textit{s} =(1,-1,1,1,-1,1,-1,1)$ corresponds to a  sign assigned to the eight corresponding
components of the Kronecker delta.

For the Bloch vector  of a qutrit,  orthogonal states in $\mathcal{H}^3$ do not correspond to opposite
points on $\mathcal{S}^7$, but to points of maximum opening angle of $\frac{2\pi}{3}$. A distribution of
points on $\mathcal{S}^7$ that represent physical states, the generalized Bloch sphere, is highly nontrivial
and the majority of points on $\mathcal{S}^7$ do not lead to any physical states (producing matrices with
negative eigenvalues). Mixed qutrit states are localized within the eight dimensional ball, though in
analogy to Bloch sphere, the generalized Bloch ball has a nontrivial structure \cite{caves,sommers}.

\subsection{Quantum and classical quasi-functions for qutrits}

A view based on local realties provide a classical interpretation of qubit or qutrit entanglement.  In this
description the directions on the Bloch sphere are interpreted as  random  local realities distributed with
a  classical distribution function. In this approach the correlations between systems $a$ and $b$ are
written as a statistical average
\begin{equation}
\label{classical}
 \mathcal{C}_{i,j}^{\;cl}= \int d\vec{n}_a  \int
d\vec{n}_b \; P_{cl}(\vec{n}_a ,\vec{n}_b)\; n^{i}_a\; n^{i}_a\,.
\end{equation}
In this formula the Bloch unit directions (local realities) $\vec{n}_a $ and $\vec{n}_b $ are  integrated
over  the  qubit or the qutrit Bloch sphere  with a weight function corresponding to a classical (positive
everywhere) probability distribution function $P_{cl}(\vec{n}_a,\vec{n}_b)$, which has uniform marginals.

For maximally entangled states of the qubit and the  qutrit the
 probability distributions and the corresponding correlations are
\begin{eqnarray}
 P_{cl}(\vec{n}_a ,\vec{n}_b)=& \frac{1}{4\pi}\delta^{(3)}(\vec{n}_a -\textit{s}\vec{n}_b)&\  \nonumber\\
\ &\Rightarrow\,\mathcal{C}_{ij}^{cl}= \frac{\textit{s}}{3}\delta_{ij},&\   \nonumber
\end{eqnarray}
\begin{eqnarray}
P_{cl}(\vec{n}_a ,\vec{n}_b)=& \frac{2}{9\pi^2}\delta^{(8)}(\vec{n}_a -\mathit{s}\vec{n}_b)&\  \nonumber\\
\ &\Rightarrow\,\mathcal{C}_{ij}^{cl}= \frac{\textit{s}}{8}\delta_{ij}\,.&\
\end{eqnarray}
Two different factors in the correlations  for the qubit and qutrit state  are due to different  solid
angles $4\pi$ for a qubit, and  $\frac{9\pi^2}{2}$ for  a qutrit. Calculations of the correlation functions
involve the following integrals
\begin{eqnarray}
 \frac{1}{4\pi} \int d \vec{n} \, n_i n_j = \frac{1}{3} \delta_{ij} &\mathrm{for} &  \mathrm{qubit}\,, \\
  \frac{2}{9\pi^2} \int d \vec{n} \, n_i n_j = \frac{1}{8} \delta_{ij} &\mathrm{for} &
  \mathrm{qutrit}\,.
\end{eqnarray}
As a result of this we see that classical correlations are $\frac{1}{3}$ and $\frac{1}{4}$ of the quantum
result. We will see in the next Section  that these two numbers will pay an essential role in the
separability problem involving mixed states.

The reason why these two classical probability distributions fail to describe quantum correlations given by
Eq. (\ref{corr2}) and Eq. (\ref{corr3}) is the fact that a local hidden variable theory based on a positive
distribution function of local realities cannot be equivalent to quantum mechanics (Bell inequality).

It is well known that in order to describe quantum correlations we have to replace the classical
distributions from Eq. (\ref{classical}) by nonlocal positive quasi-distributions or by local non-positive
quasi-distributions.  In the case of local and non-positive quantum distributions, we are dealing with
quantum quasi-distributions similar to the Glauber $P$-diagonal representation for a harmonic oscillator or
the atomic coherent states for $N$-dimensional systems. A detailed description of these quantum
quasi-distributions   for qubits can be found in \cite{malus} and \cite{ghost}.

As a result of this approach we can write  the following two quantum distribution functions with homogeneous
marginals
\begin{eqnarray}
 P_{qm}(\vec{n}_a ,\vec{n}_b)= &\frac{3}{4\pi} \delta^{(3)}(\vec{n}_a -\textit{s}\vec{n}_b)-2\left(\frac{1}{4\pi}\right)^2&\  \nonumber\\
\   &\Rightarrow\,C_{ij}= \textit{s}\delta_{ij} \,,&\  \nonumber
\end{eqnarray}

\begin{eqnarray}
P_{qm}(\vec{n}_a ,\vec{n}_b)= &\frac{8}{9\pi^2}\delta^{(8)}(\vec{n}_a -\textit{s}\vec{n}_b)
 -3\left(\frac{2}{9\pi^2} \right)^2&\nonumber\\
\  &\Rightarrow\,C_{ij}= \frac{\textit{s}}{2}\delta_{ij}\,& \label{qdistribution}
\end{eqnarray}
As we can see these two  Bloch sphere  non-positive distributions of the qubit and qutrit describe exactly
quantum correlations, making the Bell inequalities void.

\subsection{Separability of  Werner  qutrit states}

The Werner state for two qubits is a convex combination of a maximally entangled state of two qubits with a
maximally mixed state

\begin{eqnarray}
\rho_{W}&=&\frac{1-p}{4}\;\mathbb{I}_A\otimes\mathbb{I}_B+p\;|\Psi_{qubit}\rangle\langle\Psi_{qubit}|\label{werner_qb}\nonumber \\
&\  & \\
&=&\frac{1}{4}c_{\alpha\beta}^{qubit}\sigma_{\alpha}^A\otimes\sigma_{\beta}^B\,,\label{rho_c_qb}
\end{eqnarray}
where $0\leq p\leq 1$, $\alpha\,,\beta\in\{0,...,3\},\,\sigma_0\equiv\mathbb{I}$. For such a state a
necessary and sufficient condition for quantum separability condition is known.

The qutrit  Werner state is a convex combination of a maximally entangled state of two qutrits with a
maximally mixed state

\begin{eqnarray}
\rho_{W}&=&\frac{1-p}{9}\;\mathbb{I}_A\otimes\mathbb{I}_B+p\;|\Psi_{qutrit}\rangle\langle\Psi_{qutrit}|\nonumber\\
&\  &  \\
&=&\frac{1}{9}c_{\alpha\beta}^{qutrit}\lambda_{\alpha}^A\otimes\lambda_{\beta}^B\,,\label{rho_c_qt}
\end{eqnarray}
where $0\leq p\leq 1$, $\alpha\,,\beta\in\{0,...,8\},\,\lambda_0\equiv\sqrt{\frac{2}{3}}\mathbb{I}$. For
such a state a necessary and sufficient condition for quantum separability is unknown. The separability
condition of this state has been investigated in \cite{caves}, using a the $SU(3)$ Bloch form. Despite the
fundamental difference between these two states as far the mathematical criterion of separability is
concerned, we  shall study the qubit and the qutric case on equal footing using as a tool the quantum
distributions derived  in Eq. (\ref{qdistribution}) to construct the corresponding  distribution functions
for the Werner states. The density operators form a convex set, and as a result of this  the Werner
quasi-distribution functions  are convex combinations of the corresponding distributions
\begin{eqnarray}
 P_{W}(\vec{n}_a ,\vec{n}_b)&=& (1-p)\;P_{max}(\vec{n}_a
 ,\vec{n}_b)+\nonumber\\
&\  &+ p\;P_{\psi}(\vec{n}_a ,\vec{n}_b)\,,
\end{eqnarray}
where $ P_{max}(\vec{n}_a
 ,\vec{n}_b)$ is the quantum  distribution corresponding to a maximally mixed
 state. This function is different for the qubit and the qutrit
\begin{equation}
\begin{array}{ccc}
  P_{max}^{qubit}(\vec{n}_a
 ,\vec{n}_b) &=& (\frac{1}{4\pi})^2,\\
  P_{max}^{qutrit}(\vec{n}_a
 ,\vec{n}_b) &=& (\frac{2}{9\pi^2})^2\,.
\end{array}\end{equation}

The quantum  distribution corresponding to a
 maximally entangled states of the qubit or the qutrit are given by Eq. (\ref{qdistribution})
 calculated. As a result we obtain for the two Werner states

\begin{eqnarray}
 P_{W}^{qubit}(\vec{n}_a ,\vec{n}_b)&=& \frac{3p}{4\pi}\delta^{(3)}(\vec{n}_a -\textit{s}\vec{n}_b)+\nonumber\\
\  &+&\frac{(1-3p)}{(4\pi)^2}
 \nonumber \\
 P_{W}^{qutrit}(\vec{n}_a ,\vec{n}_b)&=& \frac{8p}{9\pi^2}\delta^{(8)}(\vec{n}_a
 -\mathit{s}\;\vec{n}_b)+\nonumber\\
&+&(1 -4p)(\frac{2}{(3\pi)^2})^2\,.
\end{eqnarray}
For values of $p$  for which  these distributions are positive everywhere, the mixed Werner state has a
classical interpretation and as a result is separable.  We obtain that the Werner state is separable if
$p\leq\frac{1}{3}$ for a qubit and $p\leq\frac{1}{4}$ for a qutrit.

If we let the state to evolve with time under the action of a channel, then, as a result, we obtain the time
evolution of $c_{\alpha\beta}^{state}$ coefficients (qubit or qutrit state). Or equivalently, this results
in a change of correlation matrix $\mathcal{C}_{ij}\,\rightarrow\,\mathcal{C}_{ij}(t)$. The condition on
$P_{W}^{state}(\vec{n}_a,\vec{n}_b)$ distributions to be positive everywhere can be translated into a
condition on correlation matrix. We introduce the separability function $s_{state}(t)$ for a qutrit or qubit
Werner state $\rho_W(t)$
\begin{equation}
s_{state}(t)\,\sim\,\sum_{j}\,|\mathcal{C}_{jj}(t)|.\label{sep}
\end{equation}
which yields
\begin{eqnarray}
s_{qutrit}(t)&=&\frac{1}{12}\sum_{j=1}^8\,|c^{qutrit}_{jj}|,\,\mathrm{for}\,\mathrm{qutrit},\nonumber\\
s_{qubit}(t)&=&\frac{1}{3}\sum_{j=1}^3\,|c^{qubit}_{jj}|,\,\mathrm{for}\,\mathrm{qubit}.
\end{eqnarray}
For $t=0$ we restore the previously discussed Werner states $\rho_W$, hence $s_{state}(0)=p$. In this
representation, the Werner qutrit and qubit state are separable when $s_{qutrit}(t)\leq\frac{1}{4}$ and
$s_{qubit}(t)\leq\frac{1}{3}$ respectively.

\section{3-level atoms as qutrits}

From this point of our analysis, we will consider a particular physical realization of a qutrit state,
namely 3-level atoms with energies $E_1,E_2$ and $E_3$.  Decoupling the level $|3\rangle$ from the remaining
levels, we can easily reduce the qutrit state into a qubit. There are three configuration of 3-level atoms
that can be taken into account \cite{hioe}, and we will focus on the  so called V-configuration. In the
latter, only dipole transitions depicted on \textit{Fig.1} between levels $|2\rangle \longrightarrow
|1\rangle$ and $|3\rangle \longrightarrow |1\rangle$ are allowed

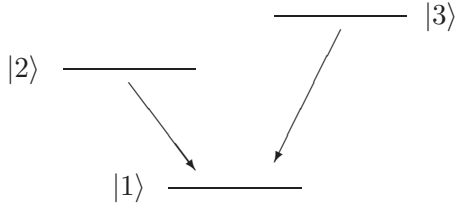
\begin{figure}[h!]
\begin{center}
\begin{picture}(200,110)
\put(140,75){\vector(-1,-2){25}} \put(60,55){\vector(3,-4){25}} \put(35,60){\line(1,0){50}}
\put(115,80){\line(1,0){50}} \put(75,15){\line(1,0){50}}

\put(20,60){\makebox(0,0){$|2\rangle$}} \put(178,80){\makebox(0,0){$|3\rangle$}}
\put(60,15){\makebox(0,0){$|1\rangle$}}

\end{picture}
\caption{Transitions allowed in 3-level atom with V-configuration}
\end{center}

\end{figure}

 Usually, the atomic
variables are populations $p_i$ and coherences $d_{ij}$ (corresponding to complex dipole moments between
states $|i\rangle$  and $|j\rangle$, with $i,j=1,2,3$). And since $\mathrm{Tr}\{ \rho\} = 1$, there are only
$8$ independent variables. These can be translated into the formalism of the qutrit Bloch vector $\vec{n}$,
namely
\begin{equation}
 \begin{array}{cccccc}
n_1 & = & \sqrt{3}\,Re\,d_{12}^*, & n_2 & = & \sqrt{3}\,\imath\,Im\,d_{12}^*,\\
n_4 & = & \sqrt{3}\,Re\,d_{13}^*, & n_5 & = & \sqrt{3}\imath\,Im\,d_{13}^*,\\
n_6 & = & \sqrt{3}\,Re\,d_{23}^*, & n_7 & = & \sqrt{3}\imath\,Im\,d_{23}^*,\\
n_3 & = & \frac{\sqrt{3}}{2}(1-2p_2-p_3),& n_8 & = & \frac{1}{2}(1-3p_3).
\end{array}\end{equation}

\subsection{Qutrit evolution in the presence of spontaneous emission}

Spontaneous emission is a dissipative process, in which the atom is coupled to electromagnetic vacuum.
Equations for the evolution of the atomic variables in presence of spontaneous emission are characterized by
decay rates, Einstein coefficients $A_2$ and $A_3$, corresponding to transitions $|2\rangle
\rightarrow |1\rangle$ and  $|3\rangle \rightarrow |1\rangle$, respectively \cite{hioe}. In a rotating frame, where
coherences $d_{ij}$ oscillate with atomic detunings $E_i-E_j$ these equations take the following form
\begin{eqnarray}
\frac{dn_{1,2}}{dt}  &=& - \frac{A_2}{2}\,n_{1,2}\,,\quad\frac{dn_{4,5}}{dt} = - \frac{A_3}{2}\,n_{4,5}\,,\nonumber\\
\frac{dn_{6,7}}{dt}  &=& - \frac{A_2+A_3}{2}\,n_{6,7}\,,
\end{eqnarray}
for the qutrit coherences and
\begin{eqnarray}
\frac{dn_{3}}{dt}  &=& - A_2 n_{3}   - \frac{\sqrt{3}}{3}( A_3 -
A_2) n_{8} +\nonumber\\
&\  &+\frac{\sqrt{3}}{6}(2 A_2 + A_3) \,,\nonumber\\
\frac{dn_{8}}{dt} &=& - A_3 n_{8}+ \frac{A_3}{2}\,,
\end{eqnarray}
for the qutrit populations. These $SU(3)$ equations for $\vec{n}(t)$  can be written in the following matrix
form
\begin{equation}
\frac{d}{dt}\vec{n}(t)=\mathcal{M}\vec{n}(0)+\vec{m}_0,
\end{equation}
where the matrix $\mathcal{M}$ and the inhomogeneous term $\vec{m}_0$ can be easily read from the previous
equations. The same equations can be expressed via the Lindblad master equation for the dissipative process
\begin{equation}
\label{lindblad}
 \frac{d\rho(t)}{dt}=\sum_k\left(
L_k\rho(t)L_k^{\dagger}-\frac{1}{2}\{\rho(t),L_k^{\dagger}L_k\}\right),
\end{equation}
with two Lindblad jump operators $L_k$
\begin{equation}
\begin{array}{ccc}
L_1 &=& \frac{1}{2}\sqrt{A_2}(\lambda_1+\imath\lambda_2),\\
L_2 &=& \frac{1}{2}\sqrt{A_3}(\lambda_4+\imath\lambda_5).
\end{array}
\end{equation}
 As a result, similar to the qubit case,  we can write the solution as
an affine transformation of the $SU(3)$ Bloch vector
\begin{equation}
\vec{n}(t) = \mathcal{D}\,\vec{n}(0) + \vec{T}(t),
\end{equation}
where the dumping matrix is
\begin{equation}
\mathcal{D}= D_{ii}  + \frac{1}{\sqrt{3}}(e^{-A_3t}-e^{-A_2t})\delta_{i3,j8}\,,
\end{equation}
with a diagonal part
\begin{eqnarray}
D_{ii}&=& (e^{-\frac{A_2t}{2}}, e^{-\frac{A_2t}{2}},e^{-A_2t},
e^{-\frac{A_3 t}{2}}, e^{-\frac{A_3 t}{2}},\nonumber\\
&\  &e^{-\frac{(A_2+A_3)t}{2}},e^{-\frac{(A_2+ A_3)t}{2}},e^{-A_3 t})\,.
\end{eqnarray}
The affine shift is a time dependent translation
\begin{eqnarray}
\vec{T}(t) &=&
\frac{1}{2\sqrt{3}}(3-e^{-A_3t}-2e^{-A_2t})\delta_{j3}+\nonumber\\
&\  &+\frac{1}{2}(1-e^{-A_3t})\delta_{j8}\,.
\end{eqnarray}
Thus,  density operator representing the state of an atom in presence of spontaneous emission is of the
form
\begin{equation}\label{rhotime}
\rho(t)=\frac{1}{3}\left(\mathbb{I}+\sqrt{3}(\mathcal{D}\,\vec{n}(0)+\vec{T}(t))\cdot\vec{\lambda}\right).
\label{rot}
\end{equation}

\subsection{Completely positive maps and Kraus opeators}

The time evolution given by Eq. (\ref{rhotime}) defines a quantum channel with noise. Any channel acting on
a density operator maps density operators into density operators \cite{daffer,ruskai,nielsen,kraus}
\begin{equation}
\Phi\ :\ \rho_{in}\mapsto \rho_{out}\,.
\end{equation}
In the case discussed in this paper  $\rho_{in}$ is the initial density operator ($\rho_{in}=\rho(0)$) and $\rho_{out}=\rho(t)$.
The interaction of a the tree-level atom with vacuum fluctuations are described by a unitary operation
acting in a Hilbert space involving the field and the atomic degrees of freedom. The reduced dynamics, if
physical, has to be described by a completely positive map that can be written in the form of the Kraus
decomposition
\begin{equation}
\rho(t)= \sum_i \mathcal{K}_i (t)\rho_{in}\mathcal{K}_i^{\dagger} (t)\,,
\end{equation}
where $\mathcal{K}_i (t) $ are time-dependent Kraus operators satisfying normalization condition:
\begin{equation}
\sum_i \mathcal{K}_i^{\dagger}(t)\mathcal{K}_i(t)=\mathbb{I}\,. \label{norm}
\end{equation}

\subsection{Kraus operators for spontaneous emission channel}

The action of spontaneous emission channel (SE channel) on the V-atom, given by equation
(\ref{rot}), can be represented by operator-sum representation. The set of Kraus operators is as follows
\begin{eqnarray}
& &\mathcal{K}_0(t) = k_{00}(t)\mathbb{I}+k_{03}(t)\lambda_3+k_{08}(t)\lambda_8,\nonumber\\
& &\mathcal{K}_1(t) = k_{11}(t)\lambda_1+k_{12}(t)\lambda_2,\nonumber\\
& &\mathcal{K}_2(t) = k_{24}(t)\lambda_4+k_{25}(t)\lambda_5,
\end{eqnarray}
where
\begin{eqnarray}
& &k_{00}(t) = \frac{1}{3}(1+e^{-\frac{A_2t}{2}}+e^{-\frac{A_3t}{2}}),\nonumber\\
& &k_{03}(t) = \frac{1}{2}(1-e^{-\frac{A_2t}{2}}),\nonumber\\
& &k_{08}(t) = \frac{1}{2\sqrt{3}}(1+e^{-\frac{A_2t}{2}}-2e^{-\frac{A_3t}{2}}),\nonumber\\
& &k_{11}(t)= \frac{1}{2}\sqrt{1-e^{-A_2t}},\nonumber\\
& &k_{12}(t) = \frac{\imath}{2}\sqrt{1-e^{-A_2t}},\nonumber\\
& &k_{24}(t) = \frac{1}{2}\sqrt{1-e^{-A_3t}},\nonumber\\
& &k_{25} (t)= \frac{\imath}{2}\sqrt{1-e^{-A_3t}}.
\end{eqnarray}
With these values of $k_{ij}(t)$ the normalization condition (\ref{norm}) is satisfied.

The time dependence of  the Kraus operators indicates that the infinitesimal $\Delta t$ evolution is
diffusive i.e. we have
\begin{eqnarray}
& &\mathcal{K}_1 (\Delta t)= \sqrt{\Delta t}L_1,\nonumber\\
& &\mathcal{K}_2 (\Delta t)= \sqrt{\Delta t}L_2,
\end{eqnarray}
where we recognize the Lindblad jump operators. As a result the dissipative evolution is equivalent to a
diffusive completely positive map that can be written in the form of the Lindblad equation (\ref{lindblad}).

\section{Influence of SE channel on state separability}

\subsection{SE channel action on qutrits}

Action of the channel produces a time dependent Werner state: $\Phi(\rho_W)=\rho_W (t)$. Therefore, the
condition $p \leq\frac{1}{4}$ to produce a separable state will be replaced by a
 time-dependent condition.
We can consider a channel that alters only one subsystem (for instance \textit{a})
\begin{equation}
\Phi_1(\rho_W)=\sum_{i=0}^2
(\mathcal{K}^a_i\otimes\mathbb{I}^b)\rho_W(\mathcal{K}^a_i\otimes\mathbb{I}^b)^{\dagger},
\end{equation}
or a channel that independently incoherently changes both subsystem
\begin{eqnarray}
& &\Phi_2(\rho_W)=q\sum_{i=0}^2 (\mathcal{K}^a_i\otimes\mathbb{I}^b)\rho_W(\mathcal{K}^a_i\otimes
\mathbb{I}^b)^{\dagger}\nonumber\\
& &+(1-q)\sum_{i=0}^2
(\mathbb{I}^a\otimes\mathcal{K}^b_i)\rho_W(\mathbb{I}^a\otimes\mathcal{K}^b)^{\dagger},
\end{eqnarray}
where  $q$ is a probability parameter (needed to satisfy (\ref{norm})). The value of $q$ is arbitrary,
though the most natural choice would be $q=\frac{1}{2}$ corresponding to a symmetric channel.

In this case the separability condition is modified and  is a function of time
\begin{eqnarray} & &s_{\mathrm{qutrit}}(t)\equiv\frac{p}{8}(e^{-A_2t}+e^{-A_3t}+2e^{-\frac{1}{2}A_2t}\nonumber\\& &
+2e^{-\frac{1}{2}A_3t}+2e^{-\frac{1}{2}(A_2+A_3)t})\leq\frac{1}{4},
\label{squtrit}
\end{eqnarray}
where for $t=0$ we have recover the initial condition
\begin{equation}
s_{\mathrm{qutrit}}(0)=p\leq \frac{1}{4}.
\end{equation}
The function $s_{\mathrm{qutrit}}(t)$ is shown on \textit{Fig.2}. We use dimensionless parameter $A_1t$
instead of \textit{t} itself, where $A_1$ is the Einstein coefficient for qubit case (it appears in qubit
channel discussion). We introduce dimensionless parameters $A_{21}\equiv\frac{A_2}{A_1}$ and
$A_{31}\equiv\frac{A_3}{A_1}$ which will illustrate the relative value of parameters $A_1,\,A_2$ and $A_3$.
Maximally entangled state (meaning $p=1$) becomes separable with time. Two cases are shown, describing the
SE channel characterized by different parameter values. Function $s_{\mathrm{qutrit}}(t)$ is symmetric with
respect to the change $A_2 \leftrightarrow A_3$, however, the values of these parameters change the time in
which $s_{\mathrm{qutrit}}(t)$ reaches the treshold value $\frac{1}{4}$. Since the maximally entangled state
loses its entanglement in a finite time, hence any less entangled state behaves in a similar way.

\begin{figure}[h!]

\includegraphics[scale=0.7]{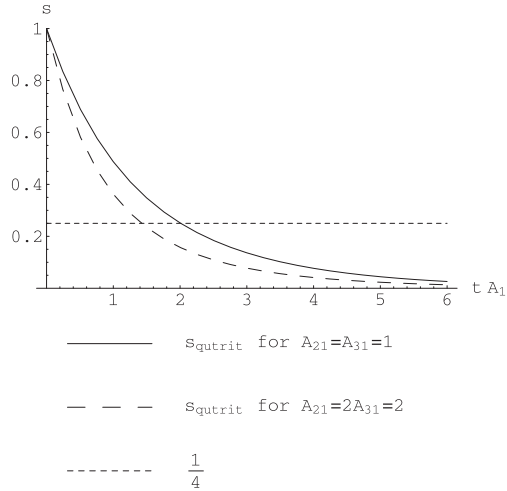}
\label{sqt} \caption{Function $s_{qutrit}(A_1t)$ for $p=1$ for two cases:
$A_{21}=\frac{A_2}{A_1}=A_{31}=\frac{A_3}{A_1}=1$ and $A_{21}=2A_{31}=2$. Region below $s=\frac{1}{4}$
corresponds to separable states}
\end{figure}

For initial pure qutrit state, $\rho_W=|\Psi\rangle\langle\Psi|$, the fidelity of the SE channel is given
by
\begin{eqnarray}
\mathcal{F}_{qutrit}(t)&=&\langle\Psi|\Phi_2(|\Psi\rangle\langle\Psi|)|\Psi\rangle,\nonumber\\
&=&\frac{1}{9}(1+e^{-\frac{A_2t}{2}}+e^{-\frac{A_3t}{2}})^2,
\end{eqnarray}
with
\begin{equation}
\mathcal{F}_{qutrit}(t\rightarrow\infty)=\frac{1}{9}\,.\label{f_qt}
\end{equation}

\subsection{SE channel and Werner state for qubits}

Kraus representation for spontaneous emission channel for qubits is given by ($A_1$ is the Einstein
coefficient)
\begin{eqnarray}
& &\mathcal{K}_0 (t)=\frac{1}{2}(1+e^{-\frac{A_1t}{2}})\mathbb{I}+\frac{1}{2}(1-e^{-\frac{A_1t}{2}})\sigma_3,\nonumber\\
& &\mathcal{K}_1 (t)=\frac{1}{2}\sqrt{1-e^{-A_1t}}(\sigma_1+\imath\sigma_2).
\end{eqnarray}

Action of the channel produces a time dependent qubit  Werner state: $\Phi(\rho_W(0))=\rho_W (t)$. Therefore,
the condition $p \leq\frac{1}{3}$ to produce a separable qubit state will be replaced by a
 time-dependent condition.

The separability condition, obtained from the same analysis as before, leads to the inequality
\begin{equation}
s_{\mathrm{qubit}}(t)\equiv\frac{p}{3}(2e^{-\frac{A_1t}{2}}+e^{-A_1t})\leq \frac{1}{3}, \label{squbit}
\end{equation}
with initial condition
\begin{equation}
s_{\mathrm{qubit}}(0)=p.\end{equation} Function $s_{\mathrm{qubit}}(t)$ is depicted on the \textit{Fig.3} as
a function of dimensionless parameter $A_1t$.
\begin{figure}[h!]
\begin{center}
\includegraphics[scale=0.7]{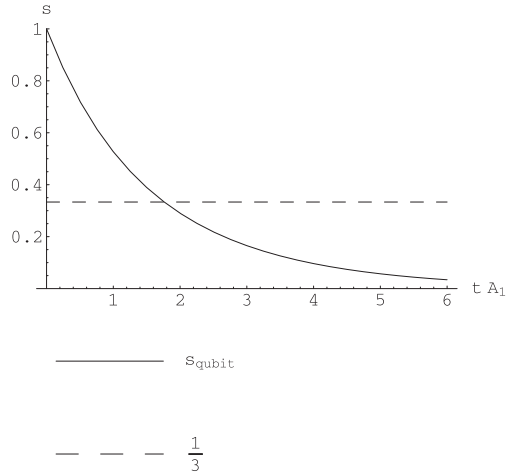}
\label{sqb} \caption{Function $s_{\mathrm{qubit}}(A_1t)$ for $p=1$. Region below $s=\frac{1}{3}$ corresponds
to separable states}
\end{center}
\end{figure}
Maximally entangled state becomes separable in a finite time, therefore any state that is less entangled becomes
separable eventually. Time in which maximally entangled state becomes separable can be calculated.\\
Assuming that the initial state is a pure qubit state, $\rho_W=|\Psi\rangle\langle\Psi|$, the fidelity of
the SE channel is given by
\begin{eqnarray}
\mathcal{F}_{qubit}(t)&=&\langle\Psi|\Phi_2^{qubit}(|\Psi\rangle\langle\Psi|)|\Psi\rangle\nonumber\\
&=&\frac{1}{4}(1+e^{-\frac{A_1t}{2}})^2.
\end{eqnarray}
with
\begin{equation}\mathcal{F}_{qubit}(t\rightarrow\infty)=\frac{1}{4}\,.\label{f_qb}
\end{equation}
\subsection{Comparison of qubit and qutrit states under the action of SE channels}

Knowing how spontaneous channel acts on both qutrit and qubit states we can compare these two cases in order
to state whether qutrit or qubit Werner states preserve entanglement longer. We show this comparison of SE
channel action on \textit{Fig.4}. Points of intersection of $s_{qutrit}$ with $\frac{1}{4}$ and $s_{qubit}$
with $\frac{1}{3}$ are crucial for this discussion. Clearly, the relative values of parameters $A_1,A_2$ and $A_3$
decide whether qutrit or qubit Werner states preserve entanglement longer.\\
From the equation (\ref{squbit}) we can read that function $s_{qubit}(t)$ reaches the treshold value
$\frac{1}{3}$ for time ($p\neq0$)
\begin{equation}
t_{qubit}= -\frac{2}{A_1}\ln (\sqrt{1+p}-1),
\end{equation}
hence for $t\ge t_{qubit}$ the qubit state $\rho(t)=\Phi_2^{qubit}(\rho_W)$ is separable. To decide whether
qutrit or qubit states are nonseparable longer, we need to evaluate $s_{qutrit}(t_{qubit})$. Qutrit
entanglement preservation is stronger if the inequality
\begin{equation}
\frac{1}{2}(\alpha^{A_{21}}+\alpha^{A_{31}})(\alpha^{A_{21}}+\alpha^{A_{31}}+2)\ge\,\frac{1}{p}
\end{equation}
is satisfied, whith $\alpha$ defined as $\alpha=\sqrt{1+\frac{1}{p}}-1$.
\begin{figure}[h!]
\begin{center}
\includegraphics[scale=0.7]{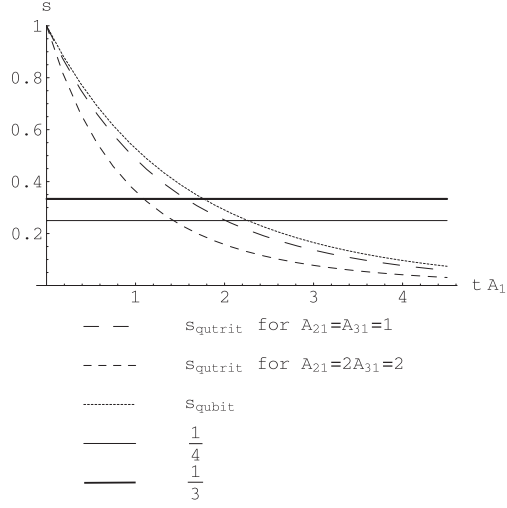}
\label{scomp} \caption{Comparison of functions $s_{qutrit}(A_1t)$ and $s_{qubit}(A_1t)$ for $p=1$ (corresponding to maximally entangled state). Points of
intersection $s_{qutrit}=\frac{1}{4}$ and $s_{qubit}=\frac{1}{4}$ show which state becomes separable first}
\end{center}
\end{figure}

In analogy, we compare fidelities of the SE channel for qubit and qutrit states. Choosing initial states to
be pure states, the fidelity values converge with time to constant values, (\ref{f_qt}) and (\ref{f_qb}).
Moreover, we can state that
\begin{equation}
\mathcal{F}_{qubit}(t\rightarrow\infty)=\mathcal{F}_{qutrit}(t\rightarrow\infty)+\frac{5}{36},\label{fid_comp}
\end{equation}
hence eventually qubit transmission through the SE channel is better. However, the separability measure is
more sensitive to relative values of parameters $A_i$ than the fidelity function. For a proper selection of
$A_i$ qutrit state  nonseparability is stronger, even though (\ref{fid_comp}) holds. The fidelity comparison
is shown on \textit{Fig.5}.

\begin{figure}[h!]

\begin{center}

\includegraphics[scale=0.8]{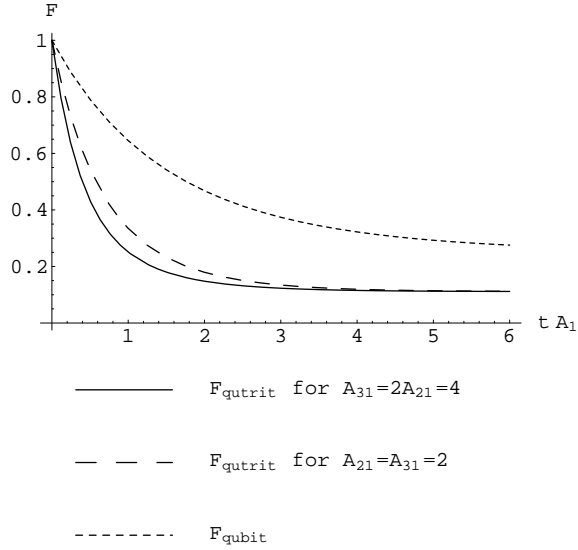}
\caption{Comparison of channel fidelities $\mathcal{F}_{qutrit}$ and $\mathcal{F}_{qubit}$ for initial pure
qutrit and qubit states} \label{fid}
\end{center}
\end{figure}

\section{Summary}

We have presented an example of a qutrit state, namely the 3-level atom with V configuration, and its
evolution under action of spontaneous emission channel. Separability of two qutrit states is, obviously,
influenced by spontaneous emission. When compared with two qubit states, Werner qutrit states may preserve
entanglement longer depending on channel parameters. This result might be of some experimental importance
when it comes to use of N-level atoms and multipartite entanglement. We plan to investigate further examples
of qutrit channels and their influence on state separability. We aim as well at general description of
qutrit channels with respect to completely positivity.

\subsection*{Acknowledgements}

This paper was supported by a MNiSW grant No. 1P03B13730. K.W. acknowledges useful discussions with
Professor C. M. Caves about entangled qutrit states.



\end{document}